\documentclass[aps,pra,reprint,showpacs,superscriptaddress]{revtex4-1} \usepackage{amsfonts}
\usepackage{graphicx,amsmath,bm,natbib,color,epstopdf}

\newcommand{\ket}[1]{|#1\rangle}
\newcommand{\bra}[1]{\langle #1|}
\newcommand{\Eq}[1]{Eq.~(#1)}

\begin{document}

\title{Process tomography via sequential measurements on a single quantum system}

\author{Humairah Bassa}
\affiliation{School of Chemistry and Physics, University of KwaZulu-Natal, Private Bag X54001, Durban 4000, South Africa}

\author{Sandeep K. Goyal}
\affiliation{School of Chemistry and Physics, University of KwaZulu-Natal, Private Bag X54001, Durban 4000, South Africa}
\affiliation{Institute of Quantum Science and Technology, University of Calgary, Alberta T2N 1N4, Canada}

\author{Sujit K. Choudhary}
\affiliation{School of Chemistry and Physics, University of KwaZulu-Natal, Private Bag X54001, Durban 4000, South Africa}
\affiliation{Institute of Physics, Sachivalaya Marg, Bhubaneswar, Odisha 751005, India}

\author{Hermann Uys}
\affiliation{Department of Physics, Stellenbosch University, Stellenbsoch 7602, South Africa}
\affiliation{National Laser Centre, Council for Scientific and Industrial Research, Pretoria 0184, South Africa}

\author{Lajos Di\'{o}si}
\affiliation{Wigner Research Center for Physics, H-1525 Budapest 114, POB 49, Hungary}

\author{Thomas Konrad}
\email{konradt@ukzn.ac.za}
\affiliation{School of Chemistry and Physics, University of KwaZulu-Natal, Private Bag X54001, Durban 4000, South Africa}
\affiliation{National Institute for Theoretical Physics (NITheP), KwaZulu-Natal 4000, South Africa}

\date{\today}

\begin{abstract}

We utilize a discrete (sequential) measurement protocol to investigate quantum process tomography of a single two-level quantum system, with an unknown initial state, undergoing Rabi oscillations. The ignorance of the dynamical parameters is encoded into a continuous-variable classical system which is coupled to the two-level quantum system via a generalized Hamiltonian. This combined estimate of the quantum state and dynamical parameters is updated by using the information obtained from sequential measurements on the quantum system  and, after a sufficient waiting period, faithful state monitoring and parameter determination is obtained. Numerical evidence is used to demonstrate the convergence of the state estimate to the true state of the hybrid system.

\end{abstract}

\pacs{03.65.Ta, 03.65.Wj, 06.20.-f}

\maketitle

\section{Introduction}

Recent years have witnessed remarkable progress in the theoretical study \cite{Diosi1988,Belavkin1989,Wiseman1993,Korotkov2000,Korotkov2001,Diosi2003,Audretsch2002,WisemanBook} and experimental realization \cite{Silberfarb2007,Haroche2011,Korotkov2012} of  the manipulation of single quantum systems in the framework of quantum information processing and communication. The ability to monitor and control single quantum systems \cite{Haroche2011,Korotkov2012,Leach2010,Wineland2011} is essential for the development of technology based on these quantum building blocks, hence methods for quantum-state monitoring and dynamical parameter estimation are of paramount importance.

A novel method for tracking the evolution of a driven, isolated two-level system in real-time by means of a sequence of measurements with minimal disturbance, was devised by Audretsch \emph{et al.}~\cite{Audretsch2001, Audretsch2007, KonradUys2012} and is related to the concept of continuous measurement \cite{Korotkov2000,Korotkov2001,Diosi2003} and state estimation \cite{Diosi2006}. The state estimate (or state guess) and the real state of the quantum system are evolved with the same propagator given by the Hamiltonian and measurement-dependent dynamics. It was argued that, for known dynamics but an unknown initial state, the estimated state and real state eventually converge~\cite{Diosi2006}. Remarkably, numerical simulations show that the convergence for  two-level systems undergoing Rabi oscillations occurs within only a few Rabi cycles \cite{KonradUys2012}. The theory also applies to many-level systems (such as a Bose-Einstein condensate in a double-well potential \cite{Hiller2012}) and systems with infinite-dimensional Hilbert spaces (for example, particles in various potentials \cite{Rothe2010}). Similar methods have been used to experimentally control the number of photons in a cavity \cite{Haroche2011}, the frequency of Rabi oscillations in a superconducting qubit \cite{Korotkov2012}, and for the state tomography of an ensemble of cold cesium atoms \cite{Silberfarb2007}.

However, these methods fail when the dynamics of the system are not precisely known -- for example, there may be some uncertainty in the Rabi frequency for an atom driven by an external laser field. Quantum process tomography, which is the task of identifying the dynamical quantum process, now becomes necessary. The standard method is extremely resource intensive since the dynamical parameters are inferred through the measurement of an informationally complete set of observables, at different times, on a large ensemble of identically prepared quantum systems \cite{Chuang1996}. As an efficient alternative, the techniques of direct characterization of quantum dynamics (DCQD) \cite{Lidar2006} and selective and efficient quantum process tomography (SEQPT) \cite{Bedensky2009} were devised. DCQD substantially reduces the number of resources required for process tomography by utilizing quantum correlations between the probe system and an ancillary qubit. SEQPT, on the other hand, achieves this objective by mapping the estimation of the dynamical parameters (specified a process matrix) to the average fidelity of the quantum channel which can, in turn, be estimated by randomly sampling over a particular set of states called a ``2''-design. However, these methods use projective measurements and therefore destroy the coherent dynamics.

In contrast, the sequential measurement protocol introduced here allows one the possibility of sensing the dynamical parameters in \emph{real time} by measuring a \emph{single} quantum system as its state evolves. The incomplete knowledge of the parameters is encoded into the state of a hypothetical classical system,  which is coupled to the quantum system via a generalized Hamiltonian. Both systems are combined to form a hybrid quantum-classical system. By means of this construction, the measurement record obtained from sequential measurements on the quantum system can be used to update the estimated hybrid state and thus the knowledge of the dynamical parameters. Essentially, we demonstrate that the problem of parameter estimation can be reduced to that of state estimation on a higher-dimensional state space. Numerical simulations demonstrate that the convergence of the estimated hybrid state to the real state is achievable in real time. It is quite remarkable that the quantum state need not be known in order to determine the dynamical parameters.

Attempts to derive an estimation scheme for the dynamical parameters via \emph{continuous} measurements were made by Negretti and M{\o}lmer \cite{Molmer2013}, Ralph \emph{et al.}~\cite{Ralph2011}, and very recently by Six \emph{et al.}~\cite{Rouchon2015}. However, they used a different estimation method and derived separate update equations for the quantum system and probability distribution of the parameters.

The remainder of the paper is structured as follows. Section \ref{Section:unsharp measurements} briefly introduces the notion of unsharp measurements as an essential tool for quantum-state estimation and monitoring. Section \ref{Section:process_tom} describes how to extend the state-estimation method for the determination of unknown parameters within the framework of the hybrid quantum-classical theory. Numerical results are analyzed in Sec.~\ref{Section:numerical} while Sec.~\ref{Section:discussion} contains a concise discussion of the estimation procedure. Section \ref{Section:conclusion} concludes.

\section{Unsharp measurements and state estimation}\label{Section:unsharp measurements}

We are interested in the monitoring, or continuous observation, of the evolution of a single quantum system in real time. Consider a two-level quantum system undergoing Rabi oscillations due to the influence of the Hamiltonian
\begin{equation}\label{Eq:H_R}
    \hat H(\Omega_R) = \frac{\hbar\Omega_R}{2}\hat{\sigma}_X,
\end{equation}
where $\Omega_R$ is the Rabi frequency and $\hat{\sigma}_X$ is the Pauli matrix that generates rotations about the $X$ axis. This system could be, for example, two levels in the hyperfine ground-state manifold of a trapped ion driven by a microwave field. The normalized state is given by
\begin{equation}
    \ket{\psi(t)} = c_0(t)\ket{0} + c_1(t)\ket{1},
\end{equation}
where $|c_0|^2 + |c_1|^2 = 1$ and $\ket{0}$ ($\ket{1}$) represents the ground (excited) state of the system. The usual way to determine the dynamics (or Rabi oscillations) of $|c_1(t)|^2$ involves the preparation of the initial state of the quantum system, followed by time-evolution of the coherent dynamics for some delay time $t$, ending with a projective measurement. This procedure is repeated for different delay times with an ensemble of quantum systems. The method can also be used when we have only a \emph{single} quantum system, but requires many repetitions of the experiment for each delay time $t$.  As an alternative, we employ an estimation method that uses a sequence of a special kind of positive-operator valued measure (POVM) measurements; so-called unsharp measurements \cite{BuschBook}, in order to gain information about the state of the system with minimal disturbance.

A POVM measurement with outcome $n$, on a system in the state $|\psi\rangle$,  will result in the postmeasurement state
\begin{equation}\label{Eq:state_change}
    |\psi_n\rangle = \frac{\hat{M}_n|\psi\rangle}{\sqrt{\langle\psi|\hat{M}_n^\dagger \hat{M}_n|\psi\rangle}}
\end{equation}
where $\hat M_n$ is the Kraus operator corresponding to the measurement result $n$. In order to monitor the dynamics of the oscillating two-level system we perform unsharp measurements of the $\sigma_z$ observable. The Kraus operators are thus given by
\begin{eqnarray}
    \hat{M}_0 &=& \sqrt{1-p_0}\ket{0}\bra{0} + \sqrt{p_0}\ket{1}\bra{1} \label{Eq:unsharp1} \\
    \hat{M}_1 &=& \sqrt{p_0}\ket{0}\bra{0} + \sqrt{1-p_0}\ket{1}\bra{1} \label{Eq:unsharp2},
\end{eqnarray}
related via $\hat{M}_0^\dagger\hat{M}_0+\hat{M}_1^\dagger\hat{M}_1= \mathbb{I}$ and $0\leq p_0\leq0.5$. The strength of the discrete measurement sequence depends on the strength of the individual measurements $\Delta p= p_0-(1-p_0)= 2p_0-1$, as well as the frequency of the measurements $1/\tau$. It is best quantified by the level-resolution rate $\gamma_m = 1/\tau_m$ with $\tau_m={\tau}/(\Delta p)^2$ \cite{Audretsch2002}.

The estimation and monitoring method of Di\'{o}si \emph{et al.}~\cite{Diosi2006} now proceeds as follows. A sequence of unsharp measurements, with a frequency $1/\tau$, is applied to the two-level system as it evolves in time according to the known Hamiltonian. The duration of the measurement is assumed to be much smaller than all other dynamical timescales (impulsive measurement approximation), therefore the state change due to measurement can still be described by Eq.~(\ref{Eq:state_change}). After $N$ measurements at time $t=N\tau$, the system is in the state
\begin{eqnarray}\label{Eq:finalstate}
    |\psi\left(N\tau\right)\rangle = \hat M_{n_N}\hat U\left(\Omega_R,\tau\right)&\hat M_{n_{N-1}}\hat U\left(\Omega_R,\tau\right)\ldots \nonumber\\
    &\times\hat M_{n_1}\hat U\left(\Omega_R,\tau\right)|\psi\rangle
\end{eqnarray}
up to an appropriate normalization constant, where
\begin{equation}\label{Eq:U_est}
    \hat U\left(\Omega_R,\tau\right)=\exp\left[-\frac{i}{\hbar}\hat H\left(\Omega_R\right)\tau\right].
\end{equation}
The same sequence of operators is then applied to a \emph{pure} estimated (or guessed) state $|\psi_e\rangle$, which is orthogonal to the real state in the worst case. It has been argued in Ref.~\cite{Diosi2006} that the effect of a sequence of measurements and measurement-independent unitary evolutions on the state of the system can be approximated in the limit of continuous measurements. In this limit the evolution is described by coupled Ito stochastic master equations for the true state of the system $|\psi\rangle$ of the system, the estimated state $|\psi_e\rangle$ and the measurement record. The analytical methods of stochastic Ito calculus were used to show the convergence of the estimated state to the real state under continuous evolution and measurement~\cite{Diosi2006}. Moreover,  the convergence of the estimated state has been tested by numerical simulations for various systems \cite{Rothe2010, KonradUys2012,Hiller2012}.

\section{Process tomography}\label{Section:process_tom}

The aforementioned method of state estimation and monitoring only works accurately if all parameters of the Hamiltonian are known precisely. Although it allows monitoring with high fidelity in the presence of small continuous noise \cite{KonradUys2012} and infrequent scattering events \cite{Rothe2010}, a lack of knowledge of any of the dynamical parameters may result in completely unfaithful outcomes.

In the following, we consider monitoring the state of a system with a single unknown dynamical parameter; for example, the Rabi frequency. In general, we know the Rabi frequency to be in some finite range, i.e., $\Omega_R \in [\omega_\mathrm{min},\omega_\mathrm{max}]$. For simplicity, we assume that there are $N$ discrete frequencies, $\omega_i$, within this range and the real frequency $\Omega_R$ is one of them. Initially, we assign an equal probability $P(\omega_i)$ {(i.e.,~an unbiased probability distribution)} to the frequencies. This corresponds to having no knowledge about which of the frequencies is the correct Rabi frequency. The first step of state estimation involves propagating the estimated state according to the Hamiltonian dynamics for a time period $\tau$.  However, under unitary time evolution [\Eq{\ref{Eq:U_est}}] with an unknown frequency, the initially pure estimated state $\ket{\psi_e}$  naturally evolves into a mixed state
\begin{equation}\label{Eq:pe_quantum}
\hat\rho_e(\tau) =\sum_{i}P(\omega_i) U(\omega_i,\tau)\ket{\psi_e}\bra{\psi_e}U^\dagger(\omega_i,\tau).
\end{equation}
We will now demonstrate how the frequencies $\omega_i$ [specified in $U(\omega_i,\tau)$] can be incorporated into an effective higher-dimensional state so that the technique of state estimation can still be applied.

We aim to formulate our estimation method for the quantum state and Rabi frequency in terms of the hybrid quantum-classical theory \cite{Aleksandrov1981, Diosi2000, Diosi2014}. For introductory purposes, we first describe the estimation technique by means of a formalism which allows us to represent the unknown dynamical parameter as the state of an additional quantum system \cite{Chase2009, Molmer2013}. In particular, we artificially construct a diagonal density matrix (i.e., a mixed \emph{quantum} state) to represent the probability distribution of the frequencies. We refer to this representation as the \emph{quantum-quantum formalism} to distinguish it from the hybrid quantum-classical formalism.  The latter reflects better the classical nature of the probability distribution as the mixed state of a classical system. Although we consider only a single unknown parameter, the result can easily be generalized to full process tomography, as shown in Sec.~\ref{Section:numerical}.

\subsection{Quantum-Quantum formalism}

The initial probability distribution of the frequencies, at time $t=0$, can be written as the state of a hypothetical quantum system as follows
\begin{equation}
   \hat\rho_\mathrm{class}(0)=\sum_{i}P(\omega_i)\ket{\omega_i}\bra{\omega_i},
\end{equation}
where $\sum_i{P(\omega_i)}=1$ and no off-diagonal  elements (coherences) are allowed for $\hat\rho_\mathrm{class}$ in this  formulation \cite{Molmer2013}. The combined estimate of the frequency and the two-level state  (at time $t=0)$ can then be expressed as the tensor product of the two corresponding states
\begin{equation}\label{Eq:total_state}
    \hat \rho^e(0) = \sum_{i}P(\omega_i)\ket{\omega_i}\bra{\omega_i}\otimes \ket{\psi_e}\bra{\psi_e}.
\end{equation}
Such a density operator acts on the Hilbert space $\mathfrak{H}_\omega\otimes\mathfrak{H}_s$ where $\mathfrak{H}_\omega$ is the Hilbert space spanned by the orthonormal vectors $\ket{\omega_i}$ corresponding to all possible dynamical parameters $\omega_i$ and $\mathfrak{H}_s$ is the Hilbert space for two-level quantum systems. We can also upgrade the Hamiltonian $\hat{H}(\Omega_R)$ [in Eq.~(\ref{Eq:H_R})] in a similar fashion and redefine it as
\begin{equation}
    \mathcal{\hat{H}} = \sum_{i}\ket{\omega_i}\bra{\omega_i}\otimes \hat H(\omega_i).
\end{equation}
The estimated state $\hat \rho^e$ [Eq.~(\ref{Eq:total_state})] evolves under this Hamiltonian as
\begin{equation}
    \hat \rho^e(\tau)=\sum_{i}P(\omega_i)|\omega_i\rangle\langle\omega_i|\otimes \hat{U}(\omega_i,\tau)|\psi_e\rangle\langle \psi_e| \hat{U}^{\dagger}(\omega_i,\tau).
\end{equation}
If we take the partial trace with respect to the first system, then we obtain the quantum state defined by Eq.~(\ref{Eq:pe_quantum}). Hence, this construction allows us to shift the ambiguity in the dynamical parameters from the Hamiltonian to the composite state. The task is now to estimate the Rabi frequency and the state of the two-level quantum system using the composite state.

The estimation experiment proceeds as previously described in Section~\ref{Section:unsharp measurements}. The quantum system evolves under the Hamiltonian $\hat H(\Omega_R)$ for a time $\tau$ after which a single unsharp measurement is performed on it, yielding measurement result $n$. The estimated composite state simultaneously evolves under $\hat{\mathcal{H}}$ for the time $\tau$ and, based on the measurement result, the augmented measurement operator
\begin{equation}
    \hat{\mathcal{M}}_n = \mathbb{I} \otimes \hat M_n.
\end{equation}
is applied on it. An unsharp measurement on the quantum system yields information not only about the true state of the system but also about the Rabi frequency, and after an evolution for time $\tau$ and a single measurement, the estimated composite state is updated as follows
\begin{align}\label{Eq:updated}
    \hat \rho^e(\tau)&\mapsto\hat \rho^{e'}(\tau) \nonumber \\
              &=\frac{1}{P(n)}\sum_{i}P(\omega_i)|\omega_i\rangle\langle\omega_i| \nonumber\\
              &\qquad\qquad\otimes \hat{M}_n \hat{U}(\omega_i,\tau)|\psi_e\rangle\langle \psi_e| \hat{U}^{\dagger}(\omega_i,\tau)\hat{M}^{\dagger}_n, \\
              &= \frac{1}{P(n)}\sum_{i}P(\omega_i)P(n|\omega_i)|\omega_i\rangle\langle\omega_i|\otimes \hat \rho_e(\omega_i,\tau),
\end{align}
where
\begin{align}
    \hat \rho_e(\omega_i,\tau)&= \frac{1}{P(n|\omega_i)}\hat{M}_n \hat{U}(\omega_i,\tau)|\psi_e\rangle\langle \psi_e| \hat{U}^{\dagger}(\omega_i,\tau)\hat{M}^{\dagger}_n,\\
    P(n|\omega_i) &= \mathrm{Tr}\left[\hat{M}_n \hat{U}(\omega_i,\tau)|\psi_e\rangle\langle \psi_e| \hat{U}^{\dagger}(\omega_i,\tau)\hat{M}^{\dagger}_n\right],\\
    P(n)           &= \sum_{i} P(\omega_i)P(n|\omega_i).
\end{align}
$P(n|\omega_i)$ is the probability of measuring result $n$ for the estimated state under the condition that it evolved through the Hamiltonian $\hat H(\omega_i) = \hbar\omega_i\hat \sigma_x/2$ up to a time $\tau$, when the measurement was performed on the system. After each measurement, the observer's knowledge about the Rabi frequency is thus updated as follows,
\begin{equation}\label{Eq:bayes}
    P(\omega_i)\to P(\omega_i|n)=\frac{1}{P(n)} P(n|\omega_i)P(\omega_i).
\end{equation}
This resembles the update of probabilities according to Bayes' Law \cite{Jaynes1990}.

As previously mentioned, after a large number $N$ of updates, the state-estimation method leads to convergence of the estimated state to the true state of the system. Since we have specified that the true Rabi frequency $\Omega_R$ is one of the frequencies $\omega_i$, the probability density will eventually approach a Kronecker-Delta function indicating that the correct frequency has been determined, i.e.,
\begin{align}
   \hat \rho^e(N\tau) &\rightarrow \sum_{i}\delta_{\Omega_R,\omega_i}\ket{\omega_i}\bra{\omega_i}\otimes\ket{\psi_e(N\tau)}\bra{\psi_e(N\tau)}\nonumber \\
    & = \ket{\Omega_R}\bra{\Omega_R}\otimes\ket{\psi_e(N\tau)}\bra{\psi_e(N\tau)}\label{Eq:guess}.
\end{align}
The estimation fidelity, which measures the overlap between the real and estimated quantum states, is exactly unity after this time, demonstrating perfect state monitoring of the single quantum system in real time, as well.

\subsection{Hybrid formalism}\label{discrete_hybrid}

We now consider estimation of the Rabi frequency and the state of a two-level quantum system within the theory of hybrid
quantum-classical systems by translating the method described in the previous section. In the hybrid formalism, the probability distribution of the frequencies, $P(\omega_i)$, can be viewed as the statistically mixed state of a hypothetical classical system while the estimate of the quantum state is treated quantum mechanically. For the estimation procedure, we construct a hybrid system, whose hybrid state (at time $t=0$) is given by
\begin{equation}\label{Eq:inital_hy}
    \widehat{\rho^e}(\omega_i,0) = P(\omega_i)\ket{\psi_e}\bra{\psi_e}.
\end{equation}
This state is positive semidefinite
\begin{equation}
    \widehat{\rho^e}(\omega_i,\tau) \geq 0, \,\,\, i = 1,2,\ldots, N
\end{equation}
and normalized
\begin{equation}
   \mathrm{Tr}\left[\sum_{i}\widehat{\rho^e}(\omega_i,\tau)\right]=1.
\end{equation}

The real state of the quantum system (at time $t=0$) can be written as
\begin{equation}
    \hat{\rho}(0)=\ket{\psi}\bra{\psi},
\end{equation}
which evolves under unitary dynamics as
\begin{equation}
    \hat{\rho}(\tau) = \hat U(\Omega_R,\tau)\hat{\rho}(0)\hat U^\dagger(\Omega_R,\tau).
\end{equation}
We prescribe the following evolution for the hybrid state, $\widehat{\rho^e}(\omega_i,\tau)$:
\begin{equation}
    \widehat{\rho^e}(\omega_i,\tau)= \hat U(\omega_i,\tau)\widehat{\rho^e}(\omega_i,0)\hat U^\dagger(\omega_i,\tau).
\end{equation}
The uncorrelated (product) structure of the hybrid state [Eq.~(\ref{Eq:inital_hy})] is immediately lost after this operation. In the hybrid formalism, the sum over the frequencies reveals the reduced state of the quantum subsystem, i.e.,
\begin{equation}
    \hat\rho_e(\tau) = \sum_{i}\widehat{\rho^e}(\omega_i,\tau),
\end{equation}
which is precisely what is specified in Eq.~(\ref{Eq:pe_quantum}). On the other hand, the trace over the hybrid state gives the reduced state of the classical system
\begin{equation}
    P(\omega_i) = \mathrm{Tr}\left[\widehat{\rho^e}(\omega_i,\tau)\right].
\end{equation}

Due to a measurement with outcome $n$, the real quantum state changes like
\begin{equation}
    \hat{\rho}(\tau) \mapsto \hat M_n \hat{\rho}(\tau) \hat M_n^\dagger
\end{equation}
and we update the estimated hybrid state as follows:
\begin{equation}
    \widehat{\rho^e}(\omega_i,\tau)\rightarrow \hat M_n\widehat{\rho^e}(\omega_i,\tau) \hat M_n^\dagger.
\end{equation}
The joint probability for the frequency $\omega_i$ and the measurement result $n$ can be obtained from the hybrid state as
\begin{equation}
    P(\omega_i,n)=\mathrm{Tr}\left[\hat M_n\widehat{\rho^e}(\omega_i,\tau)\hat M_n^\dagger\right].
\end{equation}
The updated probability distribution of the frequencies thus takes the equivalent form:
\begin{equation}
    P(\omega_i\vert n)=\frac{ P(\omega_i,n)}{\sum_{i}P(\omega_i,n)}=\frac{1}{P(n)}P(n\vert\omega_i)P(\omega_i).
\end{equation}
This result coincides completely with the quantum-quantum result, cp. Eq.~(\ref{Eq:bayes}).

\section{Numerical Simulations}\label{Section:numerical}

\subsection{Frequency estimation}

We can now test the performance of our state and parameter estimation method via numerical simulations. We want to determine the frequency of the Rabi oscillations for a two-level ion due to an external driving field as well as monitor the state of the ion. Let us assume that the Rabi frequency $\Omega_R$ is known to lie in the range [$0.95\Omega_0$,$1.05\Omega_0$] where $\Omega_0$ is an experimentally determined value. This corresponds to a relative frequency error of $5\%$ and we wish to track the Rabi frequency with a relative error of $1\%$. We therefore use a discrete grid of values in the specified range where the spacing of the values on the grid is $\Omega_0/100$, which is the required accuracy for the frequency. An unsharp measurement of a single observable, such as $\sigma_z$ in our case, on the quantum system is sufficient to determine the frequency. For our sequence of measurements, we select $\Delta p = 0.2$ as the strength of the individual measurements and a measurement period of $\tau\approx T_R/10$, where $T_R=2\pi/\Omega_R$ is the Rabi period. This selection ensures that the Rabi oscillations are only weakly disturbed due to the measurement sequence since the measurement strength $\gamma_m  \approx \Omega_R/(5\pi)$, is smaller than the Rabi frequency. This is important for monitoring the state of the ion, since a strong measurement would immediately project the ion into either the ground or excited state, i.e., it would freeze the dynamics (similar to the quantum Zeno effect~\cite{Sudarshan1977}).

Since we assume no knowledge of the actual Rabi frequency within the specified range we choose an initially flat (or unbiased) probability distribution of the frequencies while the actual probability distribution is taken to be a Kronecker-Delta function at the correct frequency. The initial quantum state estimate is chosen to be orthogonal to the real state, which is the worst case scenario. We perform 5000 measurements (or approximately 500 Rabi cycles) averaged over 1000 runs. This computation takes only 550 seconds on a desktop computer with a dual core processor. In addition, we provide, in Section~\ref{Section:discussion}, a method to approximately halve the computational time required for the determination of the frequency. Further optimization techniques will be discussed in a future work.

\begin{figure}[t!]
\centering
\includegraphics[width= \columnwidth,keepaspectratio]{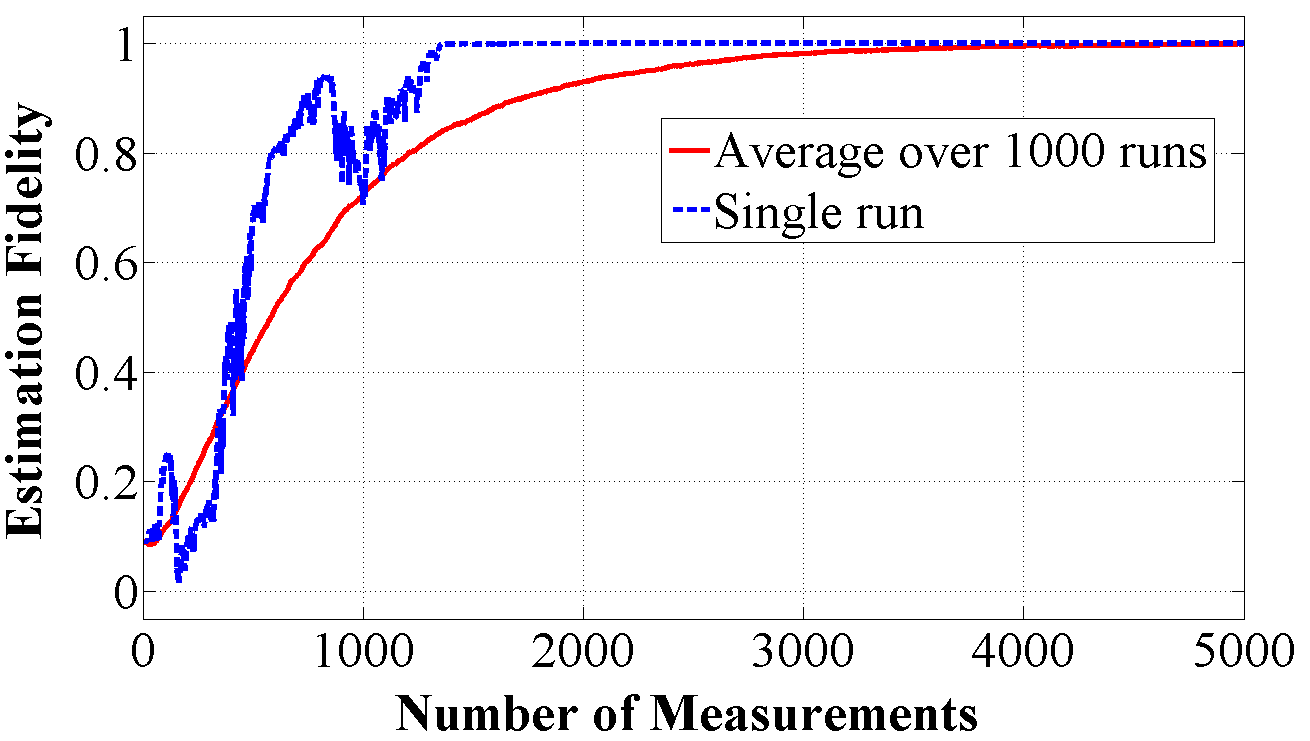}
\caption{(Color online) Estimation of the Rabi frequency. The estimation fidelity for the probability distribution of frequencies as a function of the number of measurements for a single run (dashed blue line) and averaged over 1000 runs (red line).}\label{Fig:freq}
\end{figure}

In Fig.~\ref{Fig:freq} we plot the classical estimation fidelity for the probability distribution of the frequencies as a function of the number of measurements for a single run (dashed blue line) and averaged over 1000 runs (red line). The estimation fidelity is calculated by using the well-known formula $\sum_i \sqrt{p_i q_i}$ for probability distributions $p_i$ and $q_i$. The graph asymptotically tends to unity and at this point we are able to determine the frequency within the specified accuracy. An estimation fidelity for the frequency with arbitrary precision can be obtained by decreasing the grid point distance. If the relative error with which we track the frequency is smaller than $0.1\%$ then we would also achieve perfect state monitoring once the frequency has been determined \cite{KonradUys2012}.

\subsection{Full process tomography}

Consider the case of a spin-$\tfrac{1}{2}$ particle precessing due to a magnetic field. The Hamiltonian is given by
\begin{align}
    \hat H(\Omega_L)= \frac{\hbar\Omega_L}{2} \vec{n}\cdot\hat{\vec{\sigma}} \quad \mbox{with}\,\, \vec{n}\cdot\hat{\vec{\sigma}} = \sum_i n_i\sigma_i
\end{align}
where $\Omega_L$ is the Larmor frequency, $\sigma_i$ are the Pauli matrices for the  $i=x, y, z$ axes and the normalized vector $\vec{n}$ represents the direction of the magnetic field. If the magnitude and direction of the magnetic field are not precisely known, then there will also be an uncertainty in the Larmor frequency and the axis of rotation of the particle. In order to fully determine the rotation axis, it is not possible to gain sufficient information from measurements of an observable along a single axis. We must, therefore, utilize informationally-complete (IC) unsharp measurements {of noncommuting orthogonal spin observables}. We can construct the required measurement operators from the projectors which measure the spin around the $x$, $y$ and $z$ axes, $\hat P^{i}_{\pm}=\frac{1}{2}\left(\mathbb{I}\pm\hat \sigma_{i}\right)$ with $i=x,y,z$. The measurement operators are thus given by
\begin{align}
    \hat M_0^{i}&=\frac{1}{\sqrt{3}}\left(\sqrt{1-p_0}\hat P_+^{i}+\sqrt{p_0}\hat P_-^{i}\right)  \\
    \hat M_1^{i}&=\frac{1}{\sqrt{3}}\left(\sqrt{p_0}\hat P_+^{i}+\sqrt{1-p_0}\hat P_-^{i}\right)
\end{align}
related via $\hat{M_0^x}^\dagger \hat{M_0^x}+\hat{{M}_1^{x}}^\dagger \hat{M_1^x}+\hat{M_0^{y}}^\dagger \hat{M_0^y}+\hat{M_1^{y}}^\dagger \hat{M_1^y}+\hat{M_0^{z}}^\dagger \hat{M_0^z}+\hat{M_1^{z}}^\dagger \hat{M_1^z}=\mathbb{I}$.

\begin{figure}
\includegraphics[width=\columnwidth,keepaspectratio]{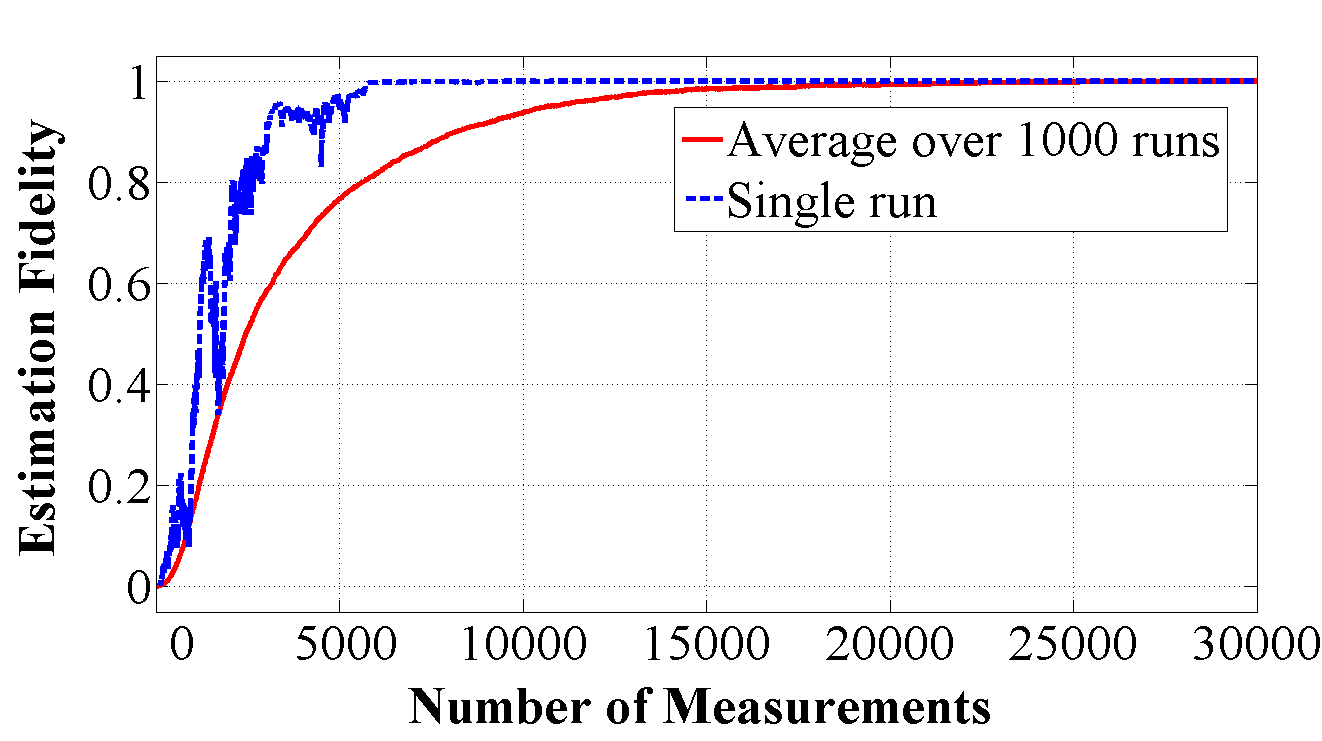}
\centering
\caption{(Color online) Estimation of the Larmor frequency and axis of rotation for a spin-$\tfrac{1}{2}$ particle. The estimation fidelity for the probability distribution of the dynamical parameters as a function of the number of measurements for a single run (dashed blue line) and averaged over 1000 runs (red line).}\label{Fig:bloch}
\end{figure}

For the simulation we consider the situation where the Larmor frequency has a relative error of $5\%$ and the axis of rotation (specified by a Bloch vector with parameters $\theta$ and $\phi$) is completely unknown. We select 10 points in the range [$0.95\Omega_0$,$1.05\Omega_0$] for experimentally determined $\Omega_0$ and 10 points each for the parameters $\theta$ and $\phi$, in the ranges $[0,\pi]$ and $[0,2\pi]$, respectively. We once again select the parameter $\Delta p = 0.2$ for the individual measurements and a measurement period of $\tau\approx T_L/10$ where $T_L=2\pi/\Omega_L$ is the Larmor period. The initial state estimate of the quantum system is taken to be orthogonal to the real state. We perform $30 000$ measurements on the quantum system and update the hybrid estimate accordingly. Figure \ref{Fig:bloch} shows the convergence of the estimated probability distribution to the actual probability distribution. We can observe that the fidelity tends, asymptotically, to unity.

\section{Discussion}\label{Section:discussion}

For our estimation scheme we utilized sequential measurements on a single quantum system instead of continuous measurements. A sequential measurement can be compared to a continuous measurement with the same measurement strength~\cite{Audretsch2002}. The measurement strength $\gamma_m$ is a specific ratio of the strength of the individual measurements $\Delta p$ and the time $\tau$ in between measurements (cp.~Section \ref{Section:unsharp measurements}). For continuous measurements the time steps $\tau$ are made infinitely small, so for the same measurement strength many more integrations per qubit cycle are required than time steps needed for sequential measurements. The estimation of the dynamical parameters of a quantum system is thus computationally more efficient with sequential measurements.

The technique of monitoring the state and frequency of the Rabi oscillations of a system via sequential measurements is characterized by two timescales \cite{Audretsch2002,Audretsch2001}: the level resolution time $\tau_m$, which defines the timescale on which the state evolves due to the measurement sequence (cp. Sec.~\ref{Section:unsharp measurements}) and $T_R$ which is the Rabi period. $\tau_m$ also characterizes the information gain due to measurement. In the case where the measurements dominate the evolution ($\tau_m \ll T_R$) the oscillations are modified and slowed down (leading to the Quantum Zeno effect for continuous projection measurements)  and the original Rabi frequency cannot be detected. On the other hand, if the measurements are very weak, i.e.,  $\tau_m\gg T_R $, the Rabi oscillations are not modified but there is little information gain per measurement and the convergence of the probability distribution of frequencies is slow. In practice, there is an optimal measurement strength that allows one to resolve the dynamical parameters with a certain accuracy in a minimum time. For example, here we used measurements with medium level-resolution time $\tau_m\approx T_R$ to resolve the Rabi frequency and the direction of the rotation axis with moderate accuracy. Optimal sequential measurements are the subject of our current and future studies, which will be presented elsewhere.

One particular advantage of our method is that the estimate or state guess can be a pure state (represented by a state vector) instead of a density matrix. We can now utilize this fact to substantially decrease the computational time required to determine the dynamical parameters. Hence, in the hybrid formalism, Section \ref{discrete_hybrid},  we replace the density matrix for the estimate state with a pure state, i.e.,
\begin{equation}
    \hat{{\rho}^e}(\omega_i) \rightarrow \sqrt{P(\omega_i)}\ket{\psi_e}.
\end{equation}
Therefore only half the number of operators are required for the estimation experiment than when density matrices are used.

\section{Conclusion}\label{Section:conclusion}

We have shown that it is possible to achieve full process tomography as well as real-time state monitoring by upgrading a known state estimation protocol. For this purpose we introduced one of the first applications of the hybrid quantum-classical formalism. Within this formalism our estimated state is a hybrid state comprised of a probability distribution of the unknown parameters and a density matrix for the quantum state. Updating the quantum part of the hybrid state according to the measurement record induces an automatic update of the probability distribution according to Bayes' Law. Numerical simulations were used to confirm the theory.

\acknowledgments{This work is based on the research supported, in part, by the National Research Foundation, South Africa (Grant specific unique reference numbers 86325 and 93602) as well as an award from the U.~S.~Air Force Office of Scientific Research (Grant No.~FA9550-14-1-0151). S.~K.~C.~acknowledges support from the Council of Scientific and Industrial Research, Government of India (Scientists' Pool Scheme). L.D. acknowledges support from the EU COST Actions MP1006, MP1209.}

\end{document}